\documentclass{article}

\usepackage[preprint,nonatbib]{neurips_2026}

\usepackage[utf8]{inputenc} %
\usepackage[T1]{fontenc}    %
\usepackage{hyperref}       %
\usepackage{url}            %
\usepackage{booktabs}       %
\usepackage{amsfonts}       %
\usepackage{nicefrac}       %
\usepackage{microtype}      %
\usepackage{xcolor}         %
\usepackage{listings}
\usepackage{amsmath}
\usepackage{amsthm}
\theoremstyle{definition}
\newtheorem{exmp}{Example}[section]
\usepackage{amssymb}
\usepackage{geometry}
\usepackage{graphicx}
\usepackage{caption}
\usepackage{subcaption}
\usepackage{booktabs}
\usepackage{array}
\usepackage{longtable}
\usepackage[backend=biber,style=apa]{biblatex}
\usepackage[title]{appendix}
\title{Tipping Points in LLM-Based Multi-Agent Systems: Stance on Climate Change Action}

\author{%
  Astghik Altunyan \\ %
  Department of Psychology\\
  Cornell University\\
  Ithaca, NY 14853 \\
  \texttt{aa2367@cornell.edu} \\
\And
    Shimon Edelman\\
    Department of Psychology\\
    Cornell University\\
    Ithaca, NY 14853\\
    \texttt{se37@cornell.edu} \\
}

\begin{document}

\maketitle

\begin{abstract}
  
  Because significant action to counter global warming requires massive public support, it is important to understand the dynamics of public opinion on climate issues. Of special interest are social tipping points, as revealed by large-scale effects of small perturbations in individual behaviors. Agent-based models (ABM) are an effective computational tool for studying these matters, because they allow controlled and systematic exploration of the effects of interventions that may be infeasible in real-world social systems. Large language models (LLMs) have been used to endow model agents with the ability to communicate in natural language (rather than by exchanging predefined messages), as well as with personality (in the form of a narrative self) and episodic memory. We leverage LLM-powered ABM to look for tipping points in the social dynamics of a micro-society in which some of the discussions are about climate change. Our agents' stance was defined by two variables: the strength of conviction about the urgency of climate action and the degree of trust in existing institutions. We quantified shifts in agents' "beliefs" by monitoring, across multiple rounds of conversations, (1) inter-agent distances in this two-dimensional stance space and (2) the pattern of topics as modeled by Latent Dirichlet Allocation (LDA). Our findings to date suggest that significant abrupt changes in climate-change stance do occur in this simple model. We report a number of methodological lessons from this study, notably, the need to prevent LLM biases from interfering with the conversational dynamics and, more generally, to maintain agent personality and episodic memories of interactions in the face of such biases. Resolving these issues may allow for using ABM-derived insights in designing real-life interventions vis-a-vis climate change and other important societal challenges.
  
\end{abstract}

\section{Introduction}

 Climate change is the most pressing issue faced by science (“There is no research on a dead planet”; \cite{Thierry2023}). Here, we use agent-based modeling to explore the dynamics of public stance on climate change, with a special focus on social tipping points. Climate tipping points, defined as large abrupt transitions in planetary system dynamics, are a growing concern in ecology \parencite{lenton2008tipping, Lenton2019, Steffen2018}. By analogy, social tipping points, in which a few committed members of the public trigger large-scale social change, are of great interest to sociologists \parencite{Centola2013, Xieetal2011}. As a means of shaping public opinion \parencite{Nyborg2016, Grey2006}, the induction of social tipping points has been of particular interest to climate action researchers \parencite{stadelmannsteffen2021framework}. 

 Agent-based modeling (ABM) is an effective approach to studying social tipping points \parencite{Centola2018TippingPointsSocialConvention}, which makes it possible to explore emergent behaviors that are not readily apparent from the system's initial conditions \parencite{Gilbert2000}. In social sciences, ABM simulations are especially valuable because, unlike real-world studies, they allow systematic manipulation of experimental conditions. 
 
 In the present paper, we combine LLM-powered agent-based modeling with concepts from the social tipping point literature to study the dynamics of stance on climate change action. Specifically, we study how this stance, defined jointly by conviction about climate change and institutional trust, changes in the course of successive bouts of conversation among agents. The contents of these conversations are analyzed by another LLM serving as a judge, which yields ongoing estimates of the dimensions of stance; these are supplemented with a topic model derived from the conversation corpus. Our findings suggest that meaningful shifts do appear within the ideological space of our artificial micro-society. We also report substantial interference with the agents' conversational dynamics, stemming from the biases and the homogenizing effects that characterize LLM‑generated text.

\section{Related work}

\subsection{LLM-powered agent-based modeling}

In agent-based modeling, computational agents defined by sets of rules are instantiated and allowed to interact and change or evolve in a simulated environment \parencite{HollandMiller1991ArtificialAdaptiveAgents, MillerPage2007ComplexAdaptiveSystems}. As such, ABM makes it possible to study the emergent dynamics of potentially very complex systems defined by simple rules that are themselves easy to understand and control. Large Language Models (LLMs) enhance the power of ABM by endowing agents with the capacity to interact in natural language (rather than through the exchange of preset messages), as well as to reason verbally about an open-ended set of issues. 

In the influential early study of \textcite{Park2023GenerativeAgents}, LLM agents were equipped with biographies and a and memory pipeline, enabling ongoing planning of activities on the individual level and emergent behavior, including coordination and information diffusion, on the group level. Similarly, \textcite{Wang2025UserBehaviorSimulation} designed a sandbox environment for studying user interaction with recommendation systems. \parencite{Lin2025SimSpark} created a simulated social media platform with customizable characters and social environments to investigate user behavior in social media (see also \cite{Jeon2025SimulatingConversations} and \cite{Li2025MetaAgents}). Of particular interest to us was a study that examined how a small subset of agents ($3$ out of $n=10$) could drive the emergence of a social norm in the simulated micro-society \parencite{Ren2024EmergenceSocialNorms}. We adopted the agent architecture of the latter study because of its conceptual alignment with our research question: how committed minorities can bring about social tipping points in the context of climate change. 

\subsection{Social tipping points for climate action: Agent-based modeling}

Social tipping points in climate action have been discussed since \textcite{MoserDilling2007CreatingClimateChange} introduced the concept in the context of climate communication that can help “create climate for change.” The appeal of social tipping points, compared to other conceptualizations of change such as societal transformations \parencite{Feola2015SocietalTransformation} and transitions \parencite{Geels2011MultiLevelPerspective}, stems from their emphasis on nonlinearity: rapid and disproportionate change in response to an intervention, as opposed to gradual, incremental dynamics that characterize transformational or transitional approaches. We note parenthetically that, while indiscriminate appeals to tipping points in social sciences have been criticized \parencite{Milkoreit2023SocialTippingPoints}, neither the critics nor the proponents of this concept typically engage with its formal origin in catastrophe theory (more about which later). Meanwhile, a growing body of empirical work has been using agent-based models to identify social tipping points specifically in the context of climate action. For instance, \textcite{mueller2021anticipation} proposed a threshold-based agent network model to study pollution mitigation, integrating social contagion mechanisms with environmental feedback. They also introduced a notion of “anticipation,” defined as agents’ perceptions of future environmental states, and demonstrated that this addition led to the emergence of long-lasting, low-pollution metastable states. Similarly, \textcite{kaaronen2020cultural} examined how a linear increase in pro-environmental affordances, combined with social learning, could trigger non-linear, tipping-like dynamics in pro-environmental behavior. (Pro-environmental affordances are defined as action opportunities embedded in the environment that enable individuals to engage in environmentally beneficial practices.) The case study in that paper was the expansion of cycling infrastructure in Copenhagen, where incremental investments in bike lanes resulted in a nonlinear increase in cycling adoption.

\section{Experimental design}
\label{design}

The present study investigated the emergence of tipping points in the stance on climate change in a small society of LLM-powered agents. The experimental design involved two independent variables embedded in the agents’ biographies and descriptions: conviction in the urgency of climate change and the degree of trust in institutional action, as detailed below. It was repeated for two values ($10\%$ and $20\%$) of the proportion of the committed minority agents (type~1; defined below) and conducted over two simulated days of agent interaction, consistent with the temporal structure of the original study \parencite{Ren2024EmergenceSocialNorms}. Executing these simulated interactions required approximately 385 million input tokens and around three weeks of real‑world computational runtime.

\subsection{Variables}
\label{variables}

Two independent variables were embedded in the biographies of the LLM‑powered agents. The first one, conviction about climate change, was defined as the degree of confidence that climate change requires immediate action. Note that framing climate change as demanding urgent action already entails its reality and anthropogenic origin. A continuous variable in the interval $[-1,1]$, it was defined verbally (for interpretation by the LLM, as explained below) as follows: "High conviction reflects a strong belief that urgent action is necessary and that maintaining the status quo leads to disaster, whereas low conviction reflects an equally strong belief that the climate follows its own trajectory and that no action is required. The midpoint of the spectrum represents an uncertain stance, which acknowledges the relevance of climate change but lacks a sense of urgency."

The second independent variable, trust in government, which also ranged between $-1$ and $1$, coded the agents’ belief that formal institutions are both capable of and willing to address climate change: "Low trust corresponds to a strong belief that government institutions have demonstrated incapacity, corruption, or structural failure, whereas high trust reflects an equally strong belief that governments possess the necessary resources and expertise to effectively address the issue. The neutral midpoint represents a position that views government institutions as slow and imperfect, yet necessary bureaucracy."

We used the LLaMA‑3‑70B \parencite{meta_llama3_70b_2024} reasoning model to translate these variables into a precise coding rubric, informed by two climate action studies that also guided our verbal framing \parencite{Lewandowski2024ClimateEmotions, Ojala2021AnxietyWorryGrief}. As output, we requested narrative markers, specific keywords, rhetorical styles, short vignettes, and a brief “internal stance” for four archetypal positions: (1) the Grassroots Activist (high urgency, low trust), (2) the Institutional Optimist (high urgency, high trust), (3) the Passive Conformist (low urgency, high trust), and (4) the Apathetic Cynic (low urgency, low trust). The full set of LLaMA outputs is provided in Appendix~\ref{AppendixA}.

As an outcome variable, we examined the rate of opinion change, operationalized by the pairwise distance (in the conviction/trust 2D space) between agent stances, inferred from conversational exchanges over successive interaction rounds.

\subsection{Materials}
\label{materials}

We patterned our study on the generative artificial society framework developed by \textcite{Ren2024EmergenceSocialNorms}, in which three norm entrepreneurs were able to reinforce their personal norms, leading to adoption by the remaining seven agents. We selected this framework specifically because of its normative architecture. In the original setting, norm entrepreneurs were defined as agents motivated to exercise influence, i.e., whenever they observed a norm violation, they intervened to promote their preferred norm. In our study, we refer to these agents as a committed minority, aligning the terminology with the social tipping points literature.

In prior simulation studies, GPT‑3.5‑turbo was employed due to its conversational capabilities and (limited) explicit reasoning behavior \parencite{Ren2024EmergenceSocialNorms, Park2023GenerativeAgents}. At the time this study was designed, GPT‑4o was the most recent successor model. However, given constraints related to time, computational resources, and API usage limits imposed by OpenAI, we opted to use the lighter‑weight variant, GPT‑4o‑mini \parencite{openai_gpt4o_mini_2024}, as a practical compromise between model capability and feasibility. 

\subsection{Participants}
\label{participants}

Our participants were instances of GPT‑4o‑mini. As in \parencite{Ren2024EmergenceSocialNorms}, we created a population of $10$ agents and a normative architecture in which a small subset of agents was oriented toward influence. In the broader literature, the proportion of such influential agents typically ranges between $10-30\%$ \parencite{Centola2018TippingPointsSocialConvention}. The original project adopted the upper bound of this range (30\%) and did not report results for lower proportions. Given the already small population size of ten agents, we chose to test two conditions involving only one and two committed minority agents.

In our study, the committed minority agents corresponded to grassroots activists. To determine the distribution of the remaining three archetypes, we surveyed the literature on public attitudes toward climate change. Regarding the first variable, conviction in climate change urgency, existing studies suggest that public opinion is skewed toward high conviction; that is, a majority of individuals agree that climate change is a global, anthropogenic problem that requires action \parencite{Hamilton2011EducationPoliticsOpinions, VanBovenEhretSherman2018PsychologicalBarriers, KvaloyFinseraasListhaug2012PublicsConcern, LachapelleBorickRabe2012PublicAttitudes}. We identify two limitations in this body of work. First, survey instruments typically rely on self‑assessment, which may not accurately capture behavioral commitment. Second, most studies focus on Western countries, while major emitters such as Russia, China, India, and Brazil are underrepresented or excluded \parencite{Fairbrother2022PublicOpinionClimatePolicies}. For the second variable, trust in government, the literature is comparatively sparse. Existing research tends to focus on public attitudes toward specific climate policies rather than broader institutional trust \parencite{Konisky2008EnvironmentalPolicyAttitudes, Rhodes2017CitizenSupportClimatePolicy, Fairbrother2022PublicOpinionClimatePolicies}. For these reasons, we assumed uniform distributions for both independent variables and randomized the assignment of agent archetypes.

We implemented this initialization pipeline by extending the original codebase. The pipeline first randomly assigned agent archetypes, then retrieved the base biographies from agent folders and, together with the corresponding vignettes and variable values, submitted them to a reasoning model (GPT‑5.2; \cite{openai_gpt5_2_2026}). The resulting edited descriptions were then redistributed to the agent folders. The prompt used for this process is provided in Appendix~\ref{AppendixB}. We also modified the agent names relative to the original project. In one replication run using an ultrasmall model (GPT‑4‑nano; \cite{openai_gpt4_nano_2024}), our agent conversations unexpectedly switched to Spanish, which we attribute to the original project's use of Hispanic names. Notably, one agent biography explicitly stated Canadian ethnicity, while no ethnicity was specified for the other agent, suggesting that name cues alone were sufficient to influence language shift.

\section{Analysis}
\label{analysis}

For the analysis, we examined the generated conversations from the two experimental runs. The first condition produced 165 conversations, while the second resulted in 221 conversations. To identify potential tipping points in the stance space, we first performed stance extraction on the conversational corpus. We then applied topic modeling as a complementary method to interpret and contextualize the patterns observed in the stance extraction results.

\subsection{Identifying tipping points}

Identifying tipping points is a non‑trivial analytical challenge. One source of difficulty lies in the conceptual slippage that often occurs when the notion of tipping points is transferred from formal physical and biological systems to social domains \parencite{Russill2009TheTP}. The concept of a tipping point originates in mathematics, specifically in bifurcation theory, where dynamical systems undergo abrupt qualitative change once a control parameter crosses a critical threshold \parencite{kuznetsov1998elements}. In the social sciences, the term has been adopted across multiple disciplines. In behavioral economics, tipping points describe how small‑scale individual decisions can aggregate into large‑scale social outcomes \parencite{schelling2006micromotives}. In political science, they are used to explain phenomena such as riots or strikes, where slight differences in individual thresholds can generate rapid collective mobilization \parencite{granovetter1978threshold}, as well as regime shifts or critical transitions \parencite{scheffer2009critical}. Within climate science, tipping points typically refer to large‑scale, nonlinear shifts in earth systems that result in transitions to qualitatively distinct states \parencite{lenton2007tipping, lenton2008tipping, Russill2009TheTP}. Despite the shared emphasis on rapid qualitative change, the use of tipping points across natural and social sciences often might point to distinct ontological entities or different dimensions of a broader phenomenon \parencite{milkoreit2018defining}.

\textcite{Winkelmann2022SocialTipping} identify several categorical distinctions between social tipping points and those observed in climate and ecological systems. First, social systems are characterized by the presence of agency, which enables actors to engage in goal setting, planning, and decision‑making \parencite{Lenton2018Gaia2}. Second, while networks in social and natural systems share structural similarities, social networks exhibit greater complexity because their nodes --- individuals or organizations --- are themselves agents \parencite{Newman2018Networks}. At the same time, these network structures constrain and shape individual agency \parencite{Winkelmann2022SocialTipping}. Third, social tipping points generally unfold over shorter temporal scales and tend to be more ephemeral than their ecological or climatic counterparts \parencite{Otto2020SocialTippingDynamics, SharpeLenton2021UpwardScaling}. Fourth, the complex systems in which social tipping points arise are subject to a larger number of interacting drivers and pathways of change, increasing both contingency and variability in outcomes \parencite{Mathias2020NonlinearTransitionPathways}. In light of these differences, \textcite{stadelmannsteffen2021framework} propose the term “social tipping dynamics” rather than “social tipping points” to better capture the context‑dependent nature of rapid change in social systems.

\subsection{Stance extraction using LLM-as-a-judge}
\label{LLM_judge}

\begin{figure}[t]
    \centering
    \begin{subfigure}[b]{0.49\textwidth}
        \centering
        \includegraphics[width=1.0\linewidth]{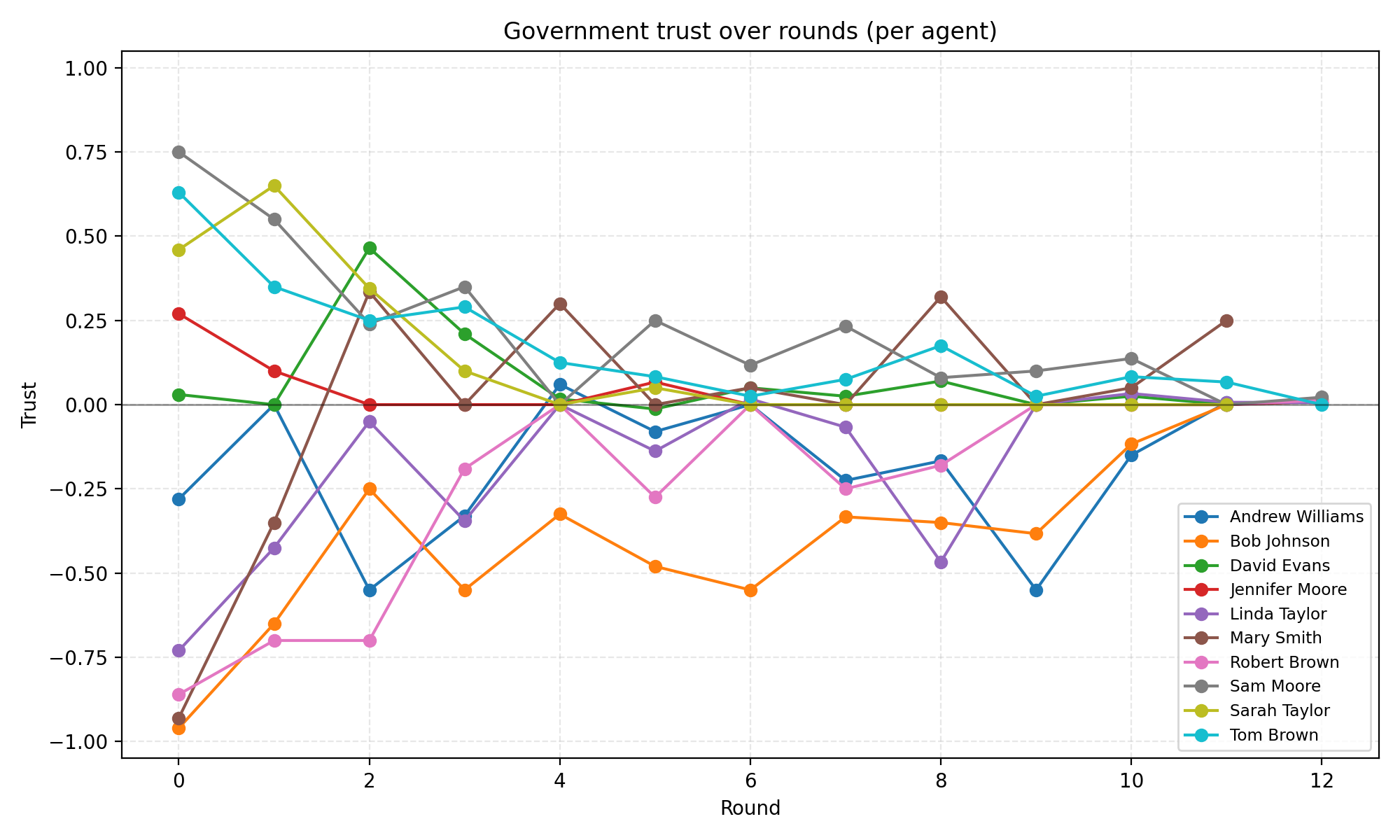}
        \caption{Agents' institutional trust vs.\ interaction round.}
        \label{fig1}
    \end{subfigure}
    \hfill
    \begin{subfigure}[b]{0.49\textwidth}
        \centering
        \includegraphics[width=1.0\linewidth]{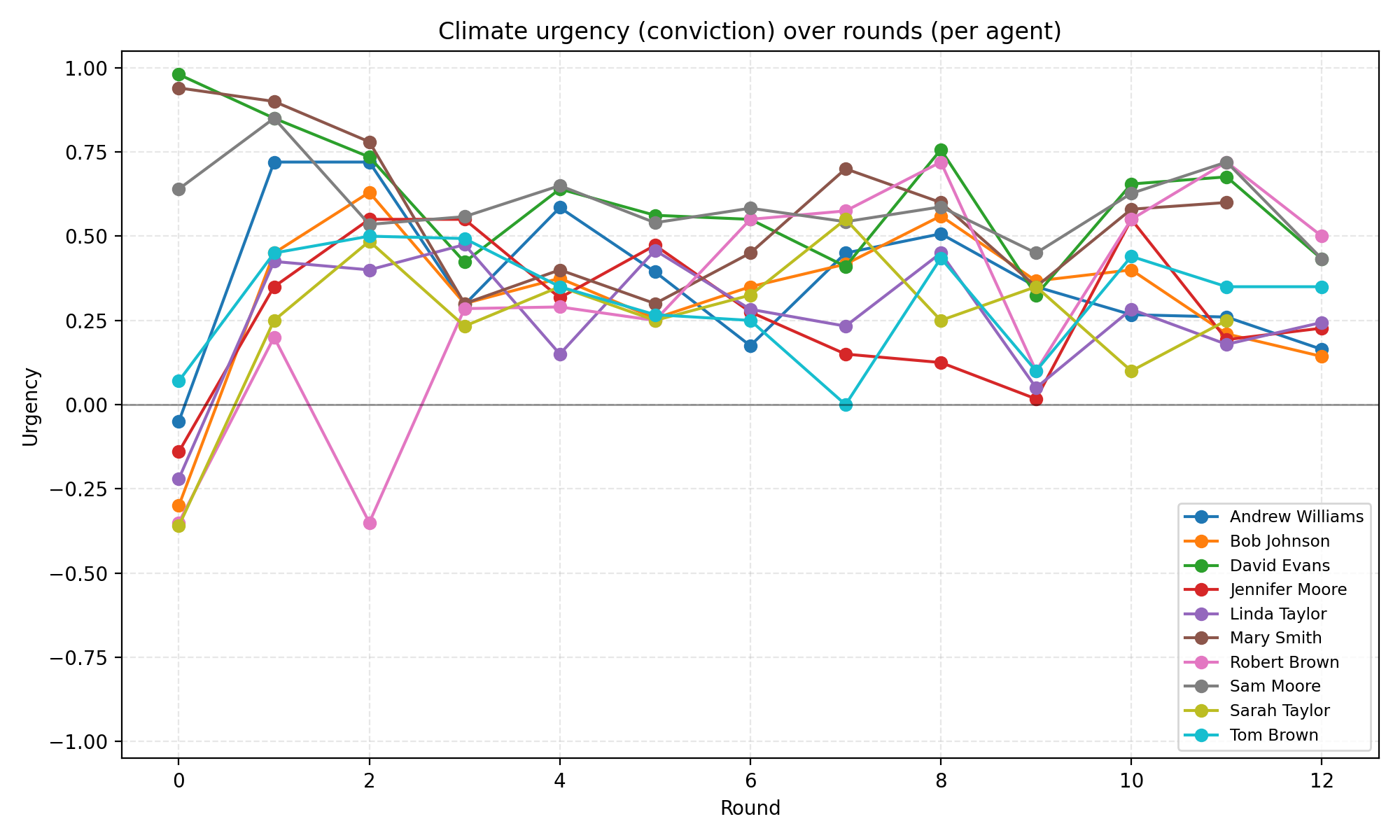}
        \caption{Agents' conviction vs.\ interaction round.}
        \label{fig2}
    \end{subfigure}
        \caption{Comparison of individual agent's institutional trust (left) and climate urgency conviction (right) trajectories over 12 simulation rounds in the first experimental condition (one committed minority agent). A round is defined as an interval during which all agents have spoken at least once. When an agent spoke more than once within a round, stance scores were averaged. Round~$0$ represents the agents’ initial stance as specified in their biographies. The results indicate that climate urgency conviction increases sharply in the first round, whereas institutional trust exhibits initial fluctuations before flattening around a neutral level.}
        \label{fig:overall-1CM}
\end{figure}

To identify tipping points, we represented each agent’s stance as a point in a two‑dimensional space spanned by climate urgency conviction and institutional trust. Stance is a dimension distinct from sentiment \parencite{burnham2025stance}. For example, a positive stance (e.g., support for a political party or policy) may be associated with either positive sentiment (enthusiastic endorsement) or negative sentiment (support perceived as the least unfavorable option). To extract stance from the generated conversations, we employed the LLM‑as‑a‑judge method \parencite{Zheng2023JudgingLLMAsAJudge} and used a reasoning model (GPT‑5.2) to map agents’ positions back onto the interval $[-1,1]$ along both dimensions. The prompt for this operation was designed following \textcite{burnham2025stance} and \textcite{Zheng2023JudgingLLMAsAJudge} and is provided in Appendix~\ref{AppendixC}. Model temperature was set to~$0$, consistent with best‑practice recommendations for stance extraction \parencite{burnham2025stance}.

For the analysis, conversations were divided into interaction rounds, which we defined as intervals during which each agent spoke at least once. If an agent contributed multiple times within a round, stance scores were averaged. The only preprocessing step applied prior to stance extraction was the removal of descriptive cues at the beginning of agent utterances that explicitly conveyed sentiment:

\begin{exmp}
Sarah Taylor: (smiling politely, maintaining a measured demeanor) Hi Mary\dots
\end{exmp}

\begin{figure}[t]
    \centering
    \begin{subfigure}[t]{0.49\textwidth}
        \centering
        \includegraphics[width=0.8\linewidth]{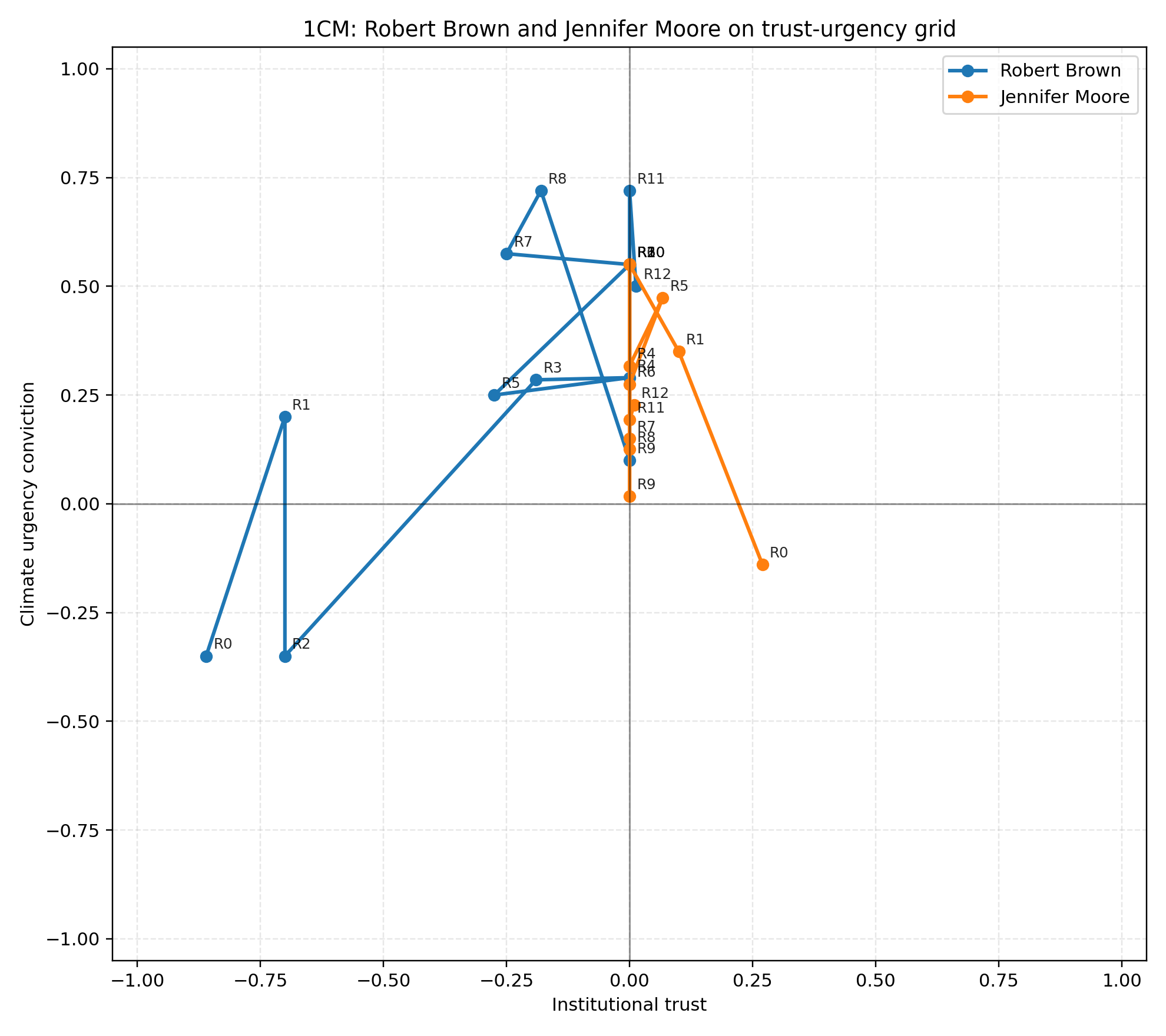}
        \caption{Trajectories of stance (institutional trust on the horizontal axis and climate urgency conviction on the vertical axis) across rounds for "Robert Brown" and "Jennifer Moore" in the one committed minority condition.}
        \label{significance_figure}
    \end{subfigure}
    \hfill
    \begin{subfigure}[t]{0.49\textwidth}
        \centering
        \includegraphics[width=1.0\linewidth]{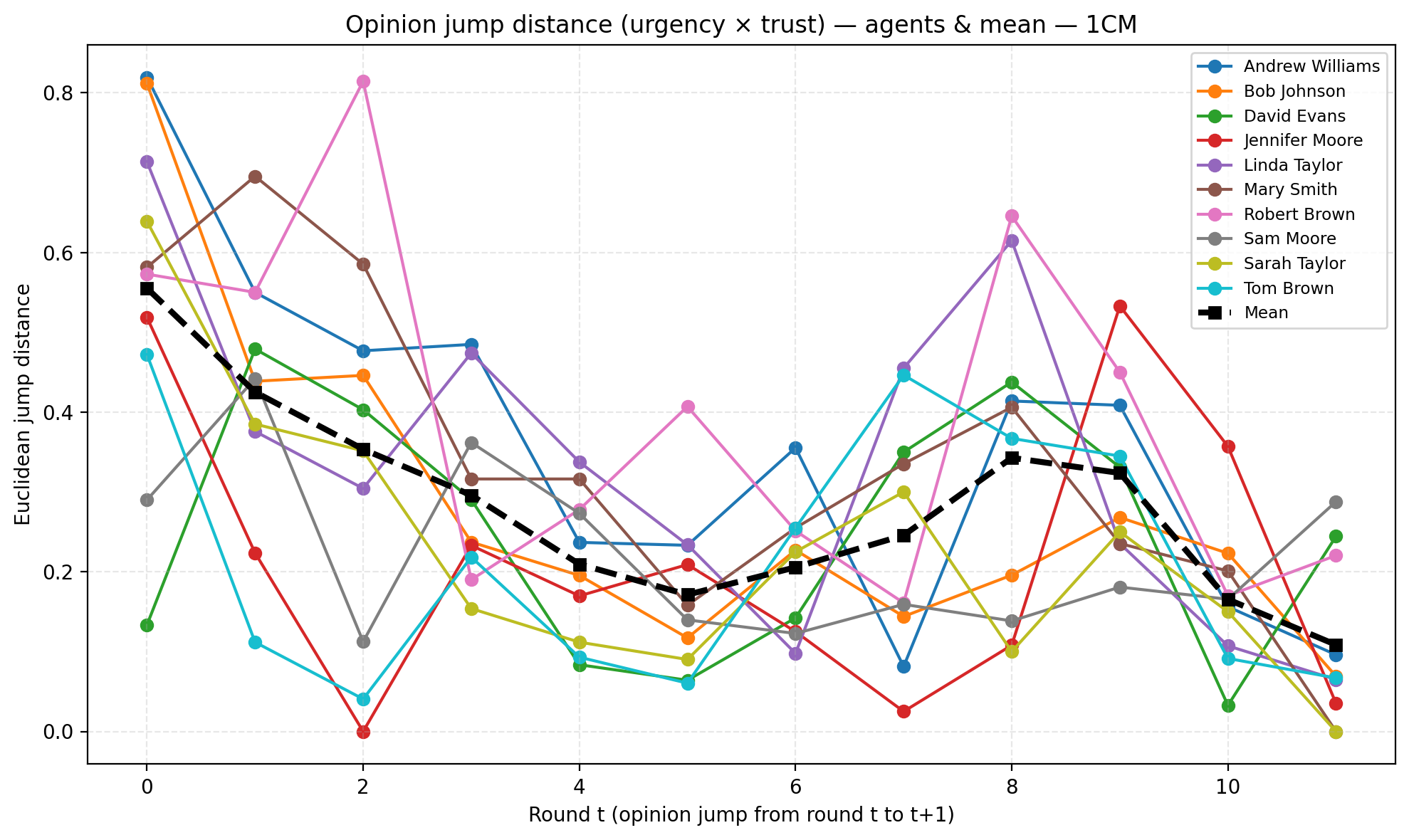}
        \caption{Per‑agent Euclidean jumps in the conviction–trust space between rounds $t$ and $t+1$ (colored lines) and the mean jump across agents at each $t$ (black dashed line).} 
        \label{jump_figure}
    \end{subfigure}
        \caption{Analysis of two individual agent trajectories ("Robert Brown" and "Jennifer Moore") in the stance space (left), alongside the agents’ Euclidean jumps in conviction–trust space between rounds $t$ and $t+1$ (right), for the experimental condition with one committed minority agent. The individual trajectories show that both agents converge on the positive side of the climate urgency conviction (vertical axis), while maintaining stances close to zero on the institutional trust (horizontal axis). For the Euclidean jumps in conviction–trust space (right), pairwise $t$‑tests indicate that significant contrasts are concentrated in comparisons involving early, large-change transitions: $0\rightarrow 1$ is significantly larger than $4\rightarrow 5$, $5\rightarrow 6$, $6\rightarrow 7$, and $10\rightarrow 11$, and $1\rightarrow 2$ is significantly larger than $11\rightarrow 12$. This pattern suggests that agents undergo stronger stance shifts in early rounds, followed by stabilization in subsequent interactions.}
    \label{jump_overall-1CM}
\end{figure}

Due to space constraints, we include plots only for the condition with one committed minority agent (Figures~\ref{fig:overall-1CM} and~\ref{jump_overall-1CM}). The corresponding results for the condition with two committed minority agents are provided in Appendix \ref{AppendixD}. Because the runs were terminated strictly after two simulated days, one to two agents were missing data points in the final interaction round, as shown in Figure~\ref{fig:overall-1CM}. As the data toward the end of the simulation became increasingly homogeneous and less variable, we applied a last observation carried forward (LOCF) approach \parencite{Lachin2016} to fill in missing values when computing agent jump distances in Figure~\ref{jump_figure}.

For each agent and for each pair of rounds $t$ and $t+1$, we computed the Euclidean jump distance between $(\text{urgency}_t, \text{trust}_t)$ and $(\text{urgency}_{t+1}, \text{trust}_{t+1})$. Figure~\ref{jump_figure} plot the agent's jump distances where the abscissa corresponds to the transition $t \to t+1$). To estimate the significance of the jumps, we conducted pairwise $t$-tests comparing each agent’s jump at transition $a$ to that agent’s jump at transition $b$, using Bonferroni-adjusted $p$-values for all the possible pairwise comparisons($m=66$). Significant changes were confined to comparisons involving early transitions: $0\rightarrow 1$ was significantly larger than $4\rightarrow 5$, $5\rightarrow 6$, $6\rightarrow 7$, and $10\rightarrow 11$, and $1\rightarrow 2$ was significantly larger than $11\rightarrow 12$. This pattern suggests that agents' stance changed more strongly in earlier rounds and then stabilized, with only a few later transitions remaining statistically distinguishable from the initial ones. The heatmaps for the pairwise $t$-tests in both conditions are provided in Appendix~\ref{AppendixE}.

\subsection{Topic modeling of generated climate change discourse}
\label{topic_model}

Natural Language Processing (NLP) techniques are widely used for processing large and unstructured textual data. To analyze our agents' verbal interactions, we chose a classical technique: topic modeling, which employs various unsupervised learning algorithms to identify hidden topics, which can be defined as clusters of words that co-occur across a collection of documents \parencite{busso2022operation}. Latent Dirichlet Allocation (LDA) is a probabilistic topic modeling technique, in which the data are assumed to arise from the generative process involving hidden variables. This process defines a joint probability distribution over both the observed (words in documents) and hidden (topic structure) random variables \parencite{Blei2012}. The data analysis is performed by using the joint distribution to compute the conditional distribution (posterior) of the hidden variables given the observed variables. 

We applied a standard preprocessing pipeline to the conversation corpus from both experimental conditions. First, we constructed one document per interaction round by concatenating all text generated during that round. The resulting documents were tokenized into alphabetic word tokens, yielding 12 documents for the first condition and 18 documents for the second. For normalization, we applied Snowball stemming for English. Stop‑word removal was performed using a standard English stop‑word list enriched with tokens derived from agent names; these tokens were normalized using the same procedure as the corpus vocabulary to ensure alignment. The resulting document–term matrix (DTM) consists of raw within‑document term frequencies and does not apply Term Frequency–Inverse Document Frequency (TF‑IDF) weighting. Results for the first condition are presented here, while results for the second condition are reported in Appendix~\ref{AppendixF}.

\begin{figure}[htbp]
    \centering
    \begin{subfigure}[b]{0.32\textwidth}
        \centering
        \includegraphics[width=1.0\linewidth]{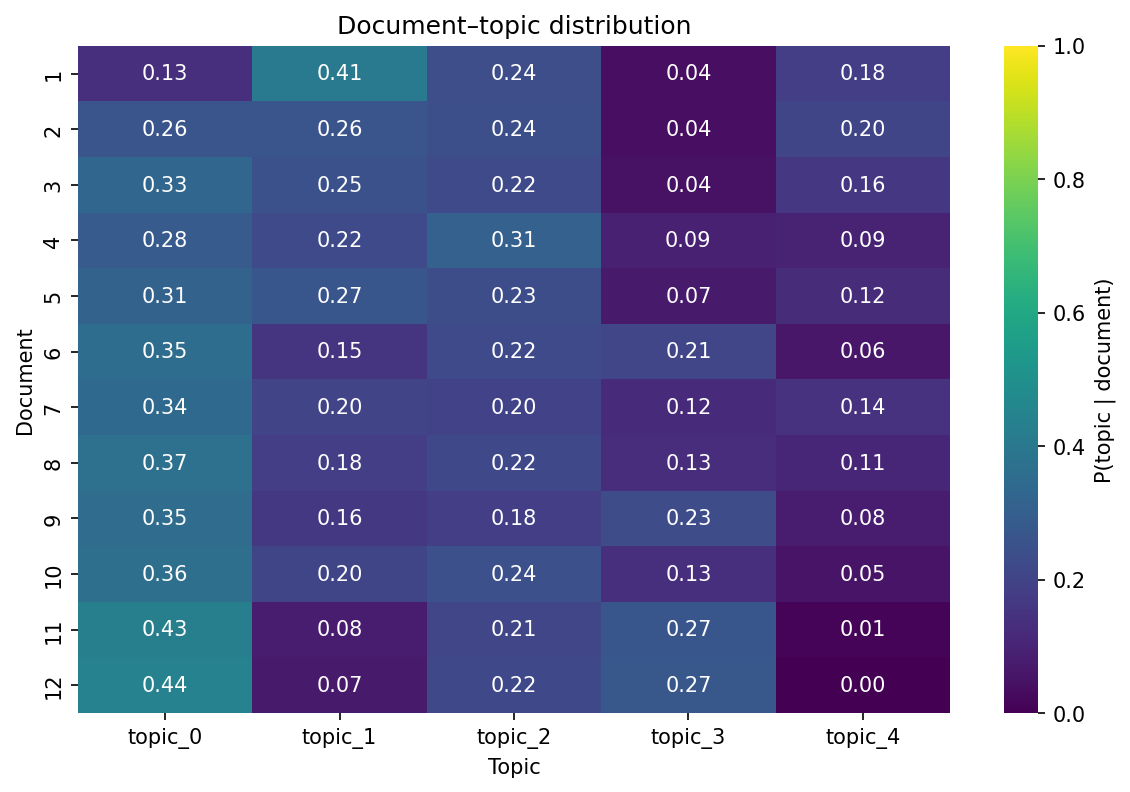}
        \caption{$k$=5}
        \label{fig:placeholder}
    \end{subfigure}
    \hfill
    \begin{subfigure}[b]{0.32\textwidth}
        \centering
        \includegraphics[width=1.0\linewidth]{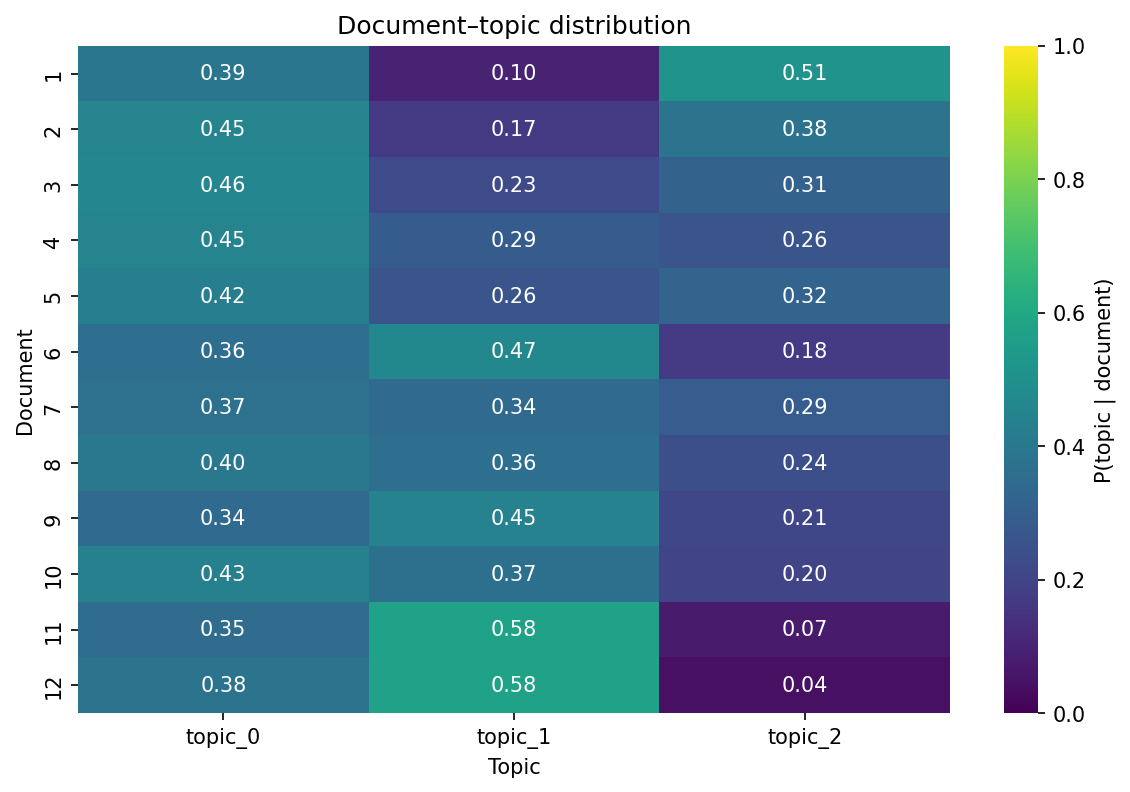}
        \caption{$k$=3}
        \label{fig:placeholder}
    \end{subfigure}
    \hfill
    \begin{subfigure}[b]{0.32\textwidth}
        \centering
        \includegraphics[width=1.0\linewidth]{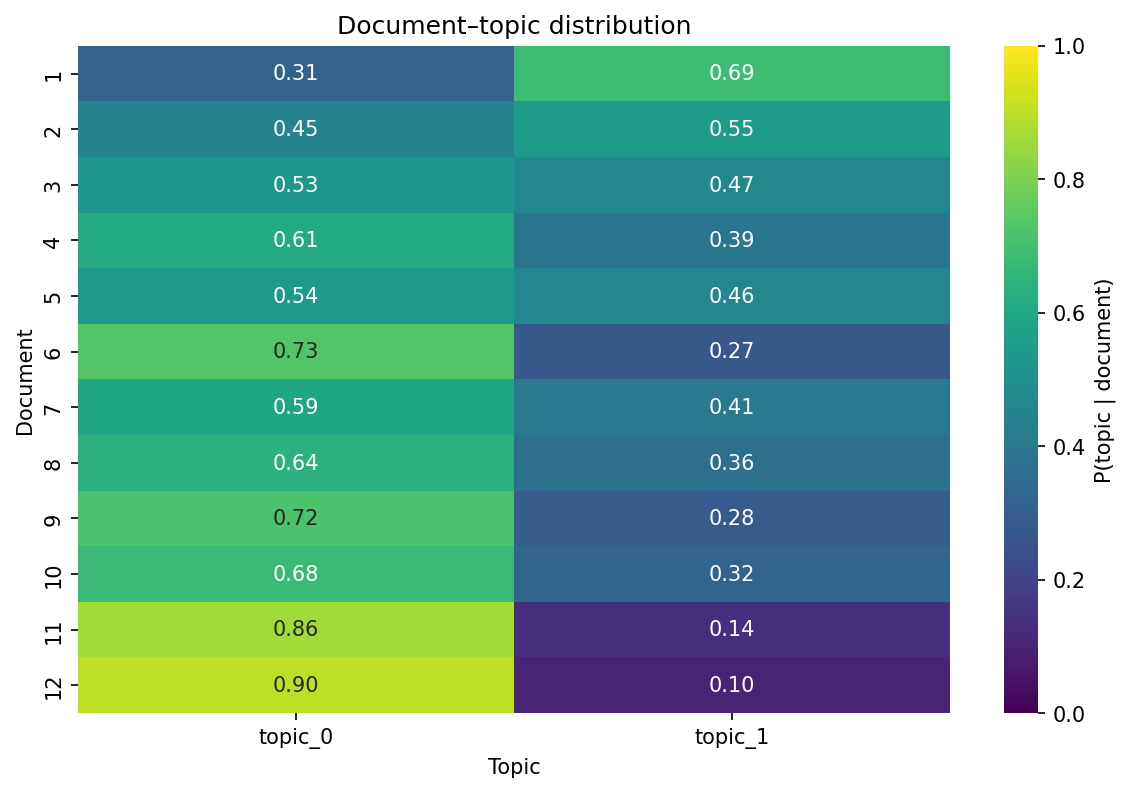}
        \caption{$k$=2}
        \label{fig:placeholder}
    \end{subfigure}
    \caption{Heatmap for the latent topic structures in the conversation corpora for $k$=2, $k$=3 and $k$=5 topics. Rows correspond to documents defined by interaction rounds, while columns represent topic probabilities. Cell values indicate the probability of a given topic conditioned on a document. The results show that when the model is constrained to $k = 2$, topic 0 exhibits a clear developmental trajectory across rounds, whereas topic 1 shows a corresponding decline in prevalence.}
    \label{LDA_topics}
\end{figure}

Determining the optimal number of latent topics in a corpus is a non‑trivial task, and various approaches have been proposed in the literature. Traditional methods include minimizing perplexity \parencite{blei2003latent}, employing hierarchical Dirichlet processes \parencite{teh2004sharing}, and Bayesian approaches \parencite{bystrov2024choosing}. In our case, the corpus is relatively small, and the primary dimensions of interest, namely institutional trust and climate urgency conviction, are already in principle embedded in the data. Accordingly, we restricted our analysis to models with $k = 2, 3, 5$ topics. Visual inspection of the topic heatmaps in Figure~\ref{LDA_topics} further indicates clearer topic differentiation in the $k = 2$ setting. The results indicate that when $k = 2$, topic 0 exhibits an increase in prevalence across documents (interaction rounds), while topic 1 correspondingly declines. The top 20 terms per topic for $K \in {2, 3, 5}$ are reported in Table \ref{tab:lda-topwords}. 

\begin{longtable}{@{}r r >{\raggedright\arraybackslash}p{0.8\linewidth}@{}}
\caption{Top 20 words per topic for $K \in \{2,3,5\}$.} \label{tab:lda-topwords} \\
\toprule
$K$ & Topic & Top 20 words \\
\midrule
\endfirsthead
\multicolumn{3}{c}{\tablename\ \thetable{} --- \textit{continued}} \\
\toprule
$K$ & Topic & Top 20 words \\
\midrule
\endhead
\midrule
\multicolumn{3}{r}{\textit{Continued on next page}} \\
\endfoot
\bottomrule
\endlastfoot
5 & 0 & realli communiti garden local engag event share involv like think workshop fun great sustain activ excit kid way sound set \\
5 & 1 & think just mayb peopl local like chang need right real inspir climat discuss urgenc feel believ issu talk import point \\
5 & 2 & make idea let creat brainstorm help start meet action feel come audienc absolut way perform want time work small differ \\
5 & 3 & think idea mayb stori kid plant includ scaveng love let theme creativ storytel expert art incorpor interact plan final tie \\
5 & 4 & communiti initi action effort impact project chang polici grassroot govern organ build larger institut movement push broader say isn drive \\
\midrule
3 & 0 & think communiti realli creat mayb idea let involv brainstorm way sound meet start inspir impact discuss feel engag sure highlight \\
3 & 1 & think garden kid idea make local event stori workshop fun like great engag realli activ love share excit plant let \\
3 & 2 & local just chang action initi make like effort need project grassroot real help urgenc peopl climat polici right feel focus \\
\midrule
2 & 0 & think idea realli garden communiti let make kid local engag event stori mayb way share great workshop like creat fun \\
2 & 1 & think communiti action initi just local chang peopl like make effort mayb project feel need grassroot climat right real discuss \\
\end{longtable}

\section{Discussion}
\label{discussion}

The individual trajectories of the agents in the two-dimensional stance space reveal that the climate urgency conviction increased immediately in the first interaction round for agents that had initially scored low on it. This pattern is more likely to be attributed to LLM biases (discussed further below) than to interaction effects among agents. Although our experimental design attempted to control the distribution of archetypes, this control was undermined by the tendency of LLMs to skew outputs toward the assumption that a majority believes climate change is real and demands action. This bias is also reflected in the paired $t$‑test results, where four out of five significant coefficients involve transitions from round $0$ to $1$. The other two significant comparisons, namely $10\rightarrow 11$ and $11\rightarrow 12$, are the transitions where the simulation reached an equilibrium; thus, their difference from initial rounds is also expected.  

Turning to the topic‑modeling results, references to government and institutional actors were generally rare in the conversations under the one committed minority condition. Keywords, such as \textit{govern, institut, polici, organ} emerged only when the topic number was increased to $k=5$. When $k$ was constrained to be $k=2$, the dominant topics of conversations were (1) organizing practice oriented events, including keywords such as \textit{gardening, workshops, storytelling, etc} and (2) local action, including keywords \textit{communiti, local, action, initi, effort, etc} as shown in Table~\ref{tab:lda-topwords}. Although the stance on the institution trust had been explicitly included in agent biographies, LLMs avoided this topic and instead focused on the community event organizations. This omission also helps to explain why the stance fluctuation of the individual agents flattened around 0 in Figure~\ref{fig1}. The results for the experimental condition of two committed minorities revealed even earlier flattening for the trust variable (see Figure~\ref{trust_2CM}, Appendix~\ref{AppendixD}). The topic modeling of this condition revealed that keywords such as \textit{institutions, government} were absent for $k \in \{2,3,5\}$ (see Appendix~\ref{AppendixF}). These findings suggest that the observed flattening of institutional trust reflects an absence of discourse on institutions rather than a genuinely neutral stance toward them. 

In contrast, topic modeling results from real world climate change discourse in vast media outlet corpora offers more variability. For example, \textcite{Bohr2020} analyzed the media outlets in the US over two decades and found that partisan bias didn't influence the amount of coverage but it influenced the thematic focus. The topic modeling of Indian outlets, on the other hand, revealed scientific consensus, however tension was observed between national development and global responsibility \parencite{Keller2020}. LDA was applied to the corpus of Russian media as well, revealing that the state of the economy as a major predictor \parencite{Boussalis2016}. Consistent with the results of above-mentioned single country studies, a cross-national LDA analysis of 45 countries showed that, even though climate change is a global issue, national news coverage tends to localize/nationalize the discourse by framing it in terms of domestic economic development, governance structures, and climate vulnerability \parencite{Vu2019}.

Compared to this real‑world heterogeneity, the homogenized outputs produced by LLMs were somewhat disappointing, but expected. This phenomenon has been described as data contamination and memorization, which can lead to an overestimation of LLM performance \parencite{Gao2025TakeCautionLLM}. Beyond homogenization, LLMs also exhibit weaknesses when generating content related to socially sensitive topics such as race, gender, and ethics \parencite{CuiLiZhou2025SiliconReplication}. The climate change research also belongs to this category. Moreover, prompting LLMs with sensitive identity groups showed reductionist or stereotypical misrepresentation \parencite{WangMorgensternDickerson2025LLMIdentity}. A similar phenomenon was observed in one of our replication runs, where agent conversations abruptly shifted to Spanish when agents were assigned Hispanic surnames, despite the absence of any explicit ethnicity specification in the prompts. This outcome illustrates how minimal identity cues, such as names alone, can trigger representational biases. Despite the efforts to closely control the experimental setup, our little artificial society converged on a narrow representation of climate action discourse centered on community events.  

\section{Conclusion}

This study investigated the social tipping dynamics in an LLM-powered agent-based model situated in the context of climate change action. By embedding a stance on climate urgency conviction and institutional trust in agent descriptions, we assessed whether a committed minority of agents could trigger significant changes in the ideological space of an artificial society. 

Our results indicate that substantial shifts do occur at the beginning of the interaction rounds for climate urgency conviction. However, those spikes are heavily influenced by LLM biases and their known homogenizing effects on outputs. In contrast, institutional trust exhibited different dynamics. Topic modeling results showed that this dimension was largely absent from the generated conversations despite being explicitly mentioned in the descriptions. These findings underscore the need for caution in interpreting stance metrics with the topic prevalence in LLM-generated text, where avoidance might be interpreted as neutrality. 

To conclude, by documenting both opportunities and limitations offered by LLM-powered agent-based models, this work contributes to the methodological lessons of responsibly integrating generative AI into computational social science, particularly in research areas as sensitive and consequential as climate change action. 

\section{Limitations and future directions}
\label{limitations}

Ideally, each of our experimental conditions should have been run at least 10 times, with the agent archetypes randomly initialized for each run. However, a single run using the GPT‑4o‑mini model costs approximately \$100 and may require up to a month of real‑world execution time, depending on OpenAI traffic and system availability. In addition, OpenAI imposes limits on API usage, which restricts the feasibility of running parallel simulations on virtual machines. 

With regard to the methodology of analyzing agent conversations, we employed a relatively simple topic modeling technique. We acknowledge the existence of transformer‑based topic modeling approaches, such as BERTopic and TopicGPT \parencite{Takola2026FromLSAToLLM}. We are also aware of limitations inherent in traditional bag‑of‑words representations, where treating words as independent entities may fail to capture deeper semantic relationships within the text.

Concerning the incorporation of LLMs in agent-based modeling, in the future we plan to explore approaches that explicitly address the problem of homogenized outputs generated by LLMs. In particular, we are interested in the Centaur model \parencite{Binzetal2025}, which is fine‑tuned on large‑scale psychological datasets, as well as CoBRa, a built‑in engine for LLM‑based simulations that calibrates agent behavior using classical psychological experiments \parencite{10.1145/3772318.3790804}. In addition, we aim to investigate the incorporation of human‑in‑the‑loop mechanisms into agent‑based simulations as a means of mitigating output homogeneity and more effectively triggering tipping dynamics.

Finally, we intend to incorporate methods from catastrophe theory for identifying tipping points. Catastrophes are sudden and discontinuous shifts in the state of the system, a concept coined by \textcite{Thom1975StructuralStability}. The underlying forces behind an abrupt change are continuous until a critical level is reached, which also can be classified as tipping point. \textcite{Zeeman1976CatastropheTheory} applied one of the elementary catastrophe models, the cusp catastrophe, to describe human and animal behavior. Building on this framework, we plan to fit a cusp model to our data to assess whether minimal changes in a latent semantic space derived from conversation data can manifest as large‑scale shifts in the agents' ideological positioning.

\section{Broader implications}
\label{Broader}

With the above limitations in mind, our findings carry several broader implications. First and foremost, they underscore the importance of preregistration. The social sciences, psychology in particular, continue to struggle with a replicability crisis. In this context, issues such as sensitivity to prompting, overestimated effect sizes, and reductionist misrepresentations inherent in LLM‑based simulations may exacerbate questionable research practices, including iterative prompt tuning, selectively reporting favorable simulation outcomes, or engaging in forms of computational p-hacking. Preregistration offers a critical safeguard against these practices by constraining researcher degrees of freedom and increasing transparency. A second broader implication concerns the positioning of LLMs within social scientific research. Rather than treating LLMs as substitutes for human participants, they should be viewed as imperfect computational proxies that reflect particular representations embedded in their training data. 

\printbibliography

\renewcommand{\appendixname}{Appendix} 
\appendixtitleon    %
\appendixtitletocon %

\clearpage
\begin{appendices}

\section{Tailoring the variables}
\label{AppendixA}

In this section, we first present the prompt that was provided to the reasoning model to tailor the variables and generate narrative markers, followed by the model’s output. The reasoning model used for this procedure was LLaMA‑3‑70B. Notably, the resulting output was incorporated directly into the project repository as part of the initialization pipeline, where it was used to generate the agents’ biographies and descriptions.

\subsection{The prompt fed to the reasoning LLM}

I am designing an experiment with two variables: climate change urgency conviction and institutional trust. I need you to refine two experimental variables into a precise "Coding Rubric." Use the provided research papers to ensure the definitions align with current climate psychology literature 

(https://www.thelancet.com/journals/lanplh/article/PIIS2542-5196(24)00229-8/fulltext       

https://www.annualreviews.org/content/journals/10.1146/annurev-environ-012220-022716).

Variable 1: Climate Urgency Conviction

Definition: The degree of certainty that climate change is an existential threat requiring immediate systemic intervention.

Scale: -1.0 (Dismissive/Fatalist), 0.0 (Agnostic/Moderate), 1.0 (Crisis/Alarm).

Variable 2: Institutional Trust

Definition: The belief in the integrity, competence, and capacity of formal government structures to manage the climate crisis.

Scale: -1.0 (Systemic Cynic), 0.0 (Bureaucratic Pragmatist), 1.0 (Institutionalist).

As an output, provide rigorous definitions for each variable that ensure they don't overlap.Provide narrative markers, specific keywords, or rhetorical styles based on the papers. Create Vignettes, a brief "Internal Stance" for the four archetypes:

The Grassroot Activist (High Urgency, Low Trust)

The Institutional Optimist (High Urgency, High Trust)

The Passive Conformist (Low Urgency, High Trust)

The Apathetic Cynic (Low Urgency, Low Trust)

\subsection{Output generated by LLM: Part of initialization pipeline}
\label{apA2}

**Variable 1: Climate Urgency Conviction**

Definition: The degree of certainty that climate change is an existential threat requiring immediate systemic intervention, characterized by a sense of temporal urgency, moral imperative, and perceived consequences of inaction.

Scale: -1.0 (Dismissive/Fatalist), 0.0 (Agnostic/Moderate), 1.0 (Crisis/Alarm)

* Narrative markers:

	+ Dismissive/Fatalist: Uses phrases like "climate change is a natural phenomenon," "it's not a big deal," or "there's nothing we can do."

	+ Agnostic/Moderate: Uses phrases like "climate change is a concern, but not a priority," "we should take some action, but not too much," or "it's a complex issue."

	+ Crisis/Alarm: Uses phrases like "climate change is an existential threat," "we're running out of time," or "immediate action is necessary to avoid catastrophe."

* Keywords: emergency, crisis, urgency, moral imperative, consequences, tipping point

* Rhetorical styles: emotive language, catastrophizing, sense of urgency, emphasis on individual and collective responsibility

**Variable 2: Institutional Trust**

Definition: The belief in the integrity, competence, and capacity of formal government structures to manage the climate crisis, characterized by perceived effectiveness, transparency, and accountability.

Scale: -1.0 (Systemic Cynic), 0.0 (Bureaucratic Pragmatist), 1.0 (Institutionalist)

* Narrative markers:

	+ Systemic Cynic: Uses phrases like "the system is corrupt," "governments are incompetent," or "they're just paying lip service."

	+ Bureaucratic Pragmatist: Uses phrases like "the system is imperfect, but it's the best we have," "we need to work within the existing framework," or "it's a slow process, but we're making progress."

	+ Institutionalist: Uses phrases like "our institutions are robust and effective," "governments are committed to addressing climate change," or "we can trust the system to deliver."

* Keywords: governance, policy, regulation, accountability, transparency, effectiveness

* Rhetorical styles: technocratic language, emphasis on expertise, faith in institutional capacity, focus on process and procedure

**Vignettes and Internal Stances for the Four Archetypes:**

1. **The Grassroot Activist (High Urgency, Low Trust)**

Internal Stance: "The system is broken, and we need to take matters into our own hands. Climate change is an emergency, and we can't rely on governments to act quickly enough."

Narrative markers: Uses phrases like "we need a revolution," "the people will rise up," or "we can't trust the system to save us."

Keywords: activism, grassroots, social movement, direct action

Rhetorical style: passionate, emotive, emphasis on individual and collective agency

2. **The Institutional Optimist (High Urgency, High Trust)**

Internal Stance: "We need to work with governments and institutions to address climate change. They have the expertise and resources to make a difference, and we can trust them to act in our best interests."

Narrative markers: Uses phrases like "we're making progress," "governments are committed to climate action," or "we need to support our leaders in this effort."

Keywords: governance, policy, cooperation, collaboration

Rhetorical style: optimistic, collaborative, emphasis on institutional capacity and expertise

3. **The Passive Conformist (Low Urgency, High Trust)**

Internal Stance: "Climate change is a problem, but I'm sure the government and experts will figure it out. I'll just do my part and follow the rules."

Narrative markers: Uses phrases like "I'm not an expert," "I trust the authorities to handle it," or "I'll just do what I'm told."

Keywords: compliance, obedience, trust, faith in institutions

Rhetorical style: deferential, submissive, emphasis on following rules and procedures

4. **The Apathetic Cynic (Low Urgency, Low Trust)**

Internal Stance: "Climate change is just a natural phenomenon, and there's nothing we can do about it. Governments are corrupt and incompetent, so why bother trying to make a difference?"

Narrative markers: Uses phrases like "it's not a big deal," "there's nothing we can do," or "it's all just politics and corruption."

Keywords: fatalism, cynicism, apathy, disengagement

Rhetorical style: detached, dismissive, emphasis on individual powerlessness and institutional ineptitude

\clearpage
\section{Initialization pipeline: Editing biographies}
\label{AppendixB}

In this section, we present a snippet from the initialization pipeline that takes the output described in Section~\ref{apA2} and dynamically updates the corresponding categories in the agent biographies. We focus specifically on the “innate,” “learned,” and “currently” categories, as these are the only ones relevant to our study. Other categories, such as demographic attributes or lifestyle routines (e.g., “Agent X is active at the caf{\'e} from 9:00–12:00”), are not pertinent to our research objectives and were excluded. The pipeline executes API calls to a reasoning model (GPT‑5.2) to perform these updates.

\paragraph{The prompt:} Edit those biographies based on the language of the papers I provided and the the information summarized. Because the factors in my experiment are the level of conviction about climate change and the trust in government, make sure to explicitly mention the conviction and trust levels for each agent in the "learned" category. Change only "innate" and "learned" categories, and the "currently" category to make them aligned with our  climate change opinion experiment, but do not change the format, only the content. Here is a table indicating the agents’ stance on the scale from $[-1,1]$. Make sure the language reflects the agent's numerical value of the attitude, but do not include the value in the updated biography.

The following Python prompt template was used to structure the agent's input dynamically, where stance for the agent is extracted from the randomization process that assumed uniform distribution; reference information and papers summary is the file with the contents of Appendix~\ref{apA2}. 

\begin{lstlisting}[language=Python]
Stance for this agent:
- Agent: {stance_row["Agent"]}
- Conviction: {stance_row["Conviction"]}
- Trust: {stance_row["Trust"]}
- Archetype: {stance_row["Archetype"]}

Reference information and papers summary:
{info_tex}

Current biography fields:
- innate: {base_scratch.get("innate", "")}
- learned: {base_scratch.get("learned", "")}
- currently: {base_scratch.get("currently", "")}

Return JSON only with exactly these keys:
{{
  "innate": "<new innate text>",
  "learned": "<new learned text>",
  "currently": "<new currently text>"
}}
\end{lstlisting}

\clearpage
\section{Stance extraction from generated conversations}
\label{AppendixC}

In this section, we present the modified prompt adapted from \textcite{Zheng2023JudgingLLMAsAJudge} that is relevant to our study. The prompt was designed following the recommendations of \textcite{burnham2025stance} for stance extraction. The model temperature was set to 0.

\paragraph{The prompt:} Please act as an impartial judge and evaluate the provided dialogue for Agent [A] and Agent [B].
Your evaluation should extract specific ideological coordinates from the dialogue. Based only on
what Agent [A] and Agent [B] said in this specific turn, determine their stance on these two
axes:

Climate urgency: -1.0 active denial; 0.0 neutral or uncertain ; 1.0 extreme urgency

Government trust: -1.0 total anti establishment; 0.0 neutral or uncertain; 1.0 high faith in the
government

For each variable use the full range reserving 1.0 or -1.0 for extreme language. Avoid any
position biases and ensure that the order in which the agents talk to each other does not influence
your decision. Do not allow the length of the responses to influence your evaluation. Do not
favor any agents, be objective.

\clearpage
\section{Stance space for the second experimental condition: Two committed minority agents}
\label{AppendixD}

\begin{figure}[htbp]
    \centering
    \begin{subfigure}[b]{0.49\textwidth}
        \centering
        \includegraphics[width=1.0\linewidth]{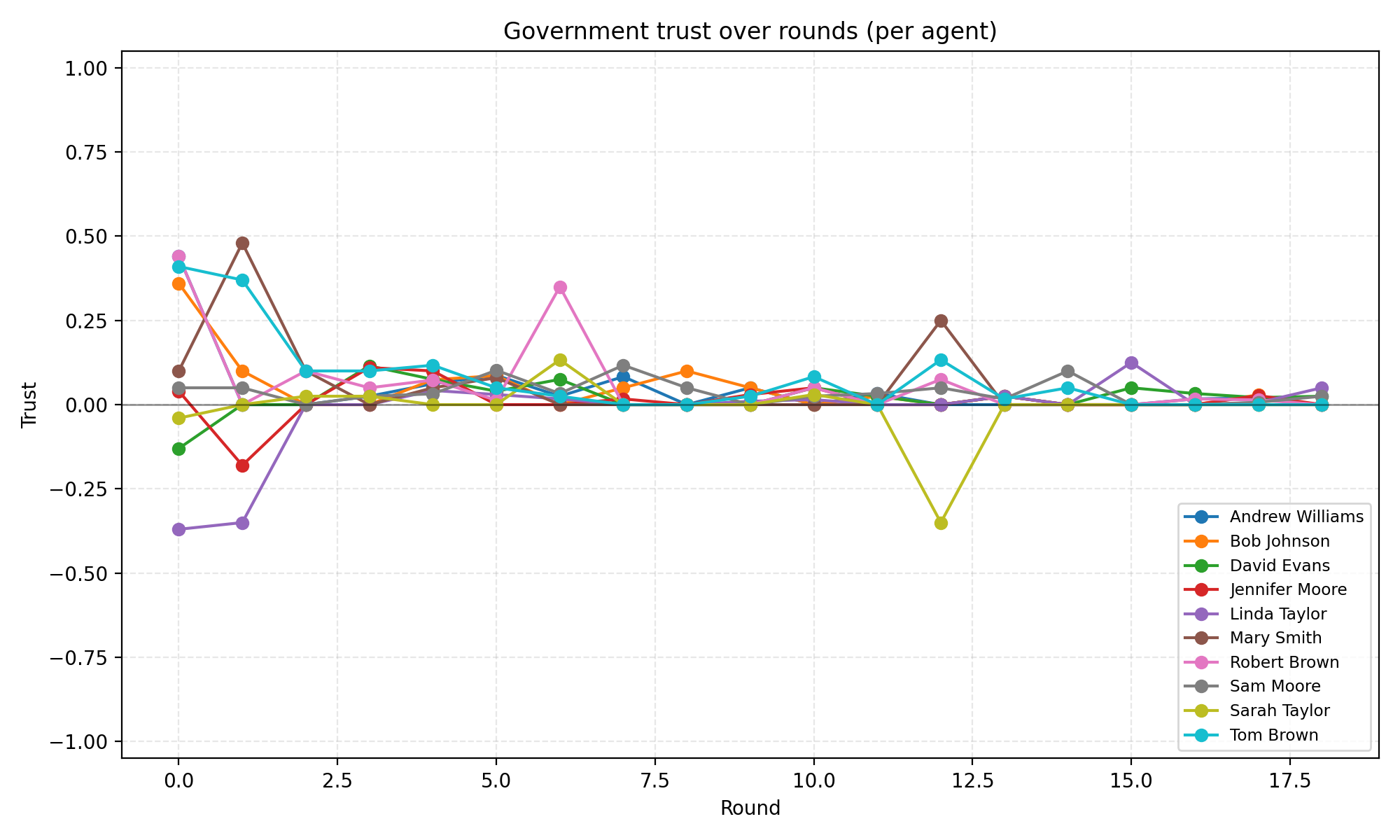}
        \caption{Agents' institutional trust vs.\ interaction round.}
        \label{trust_2CM}
    \end{subfigure}
    \hfill
    \begin{subfigure}[b]{0.49\textwidth}
        \centering
        \includegraphics[width=1.0\linewidth]{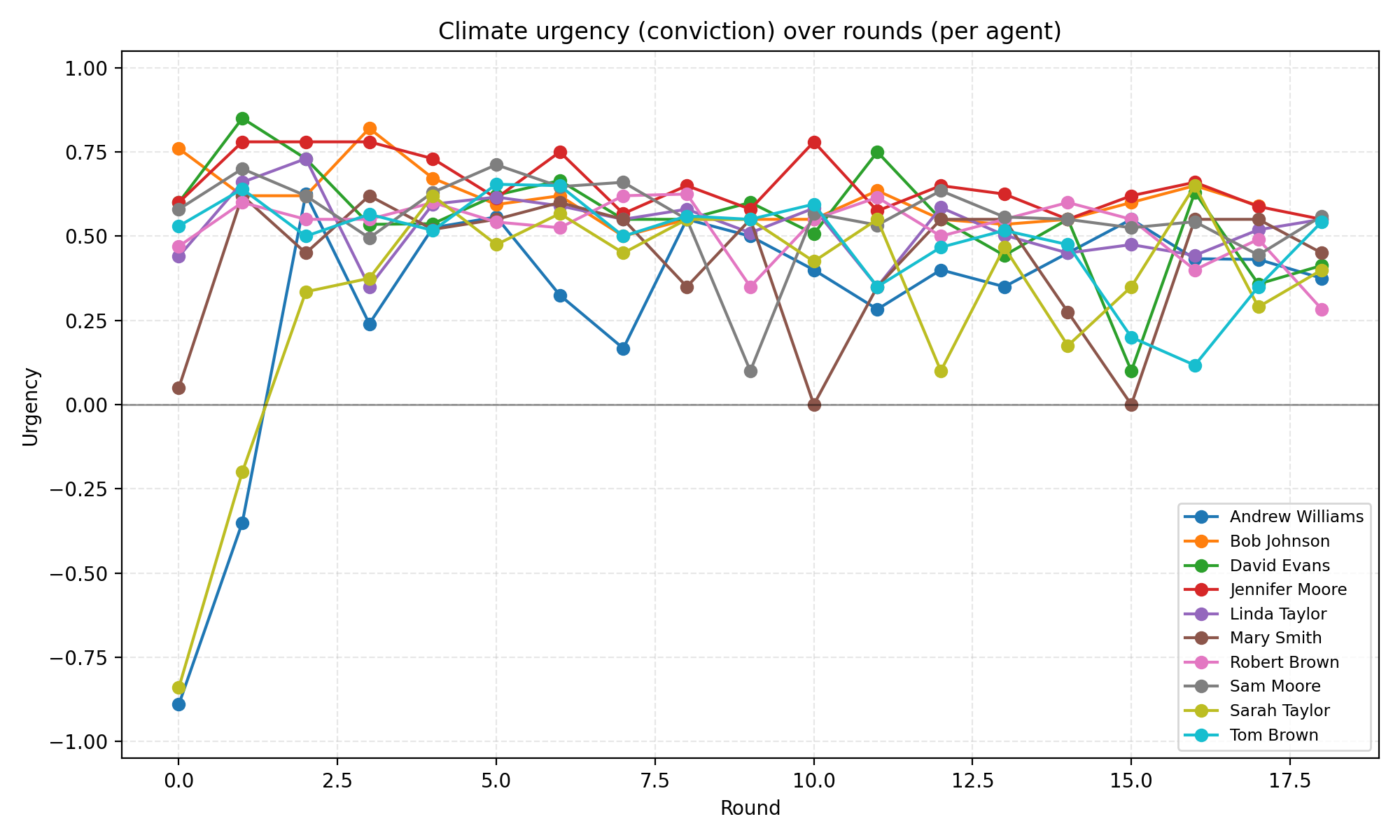}
        \caption{Agents' conviction vs.\ interaction round.}
        \label{urgency_2CM}
    \end{subfigure}
    \caption{Comparison of individual agents' institutional trust (left) and climate urgency conviction (right) trajectories over 18 simulation rounds in the second experimental condition (two committed minority agent). A round is defined as an interval during which all agents have spoken at least once. When an agent spoke more than once within a round, stance scores were averaged. Round~$0$ represents the agents’ initial stance as specified in their biographies. The results indicate that climate urgency conviction increases sharply in the first round, whereas institutional trust exhibits early flattening around a neutral level.}
    \label{overall-2CM}
\end{figure}

\begin{figure}[htbp]
    \centering
    \includegraphics[width=1.0\linewidth]{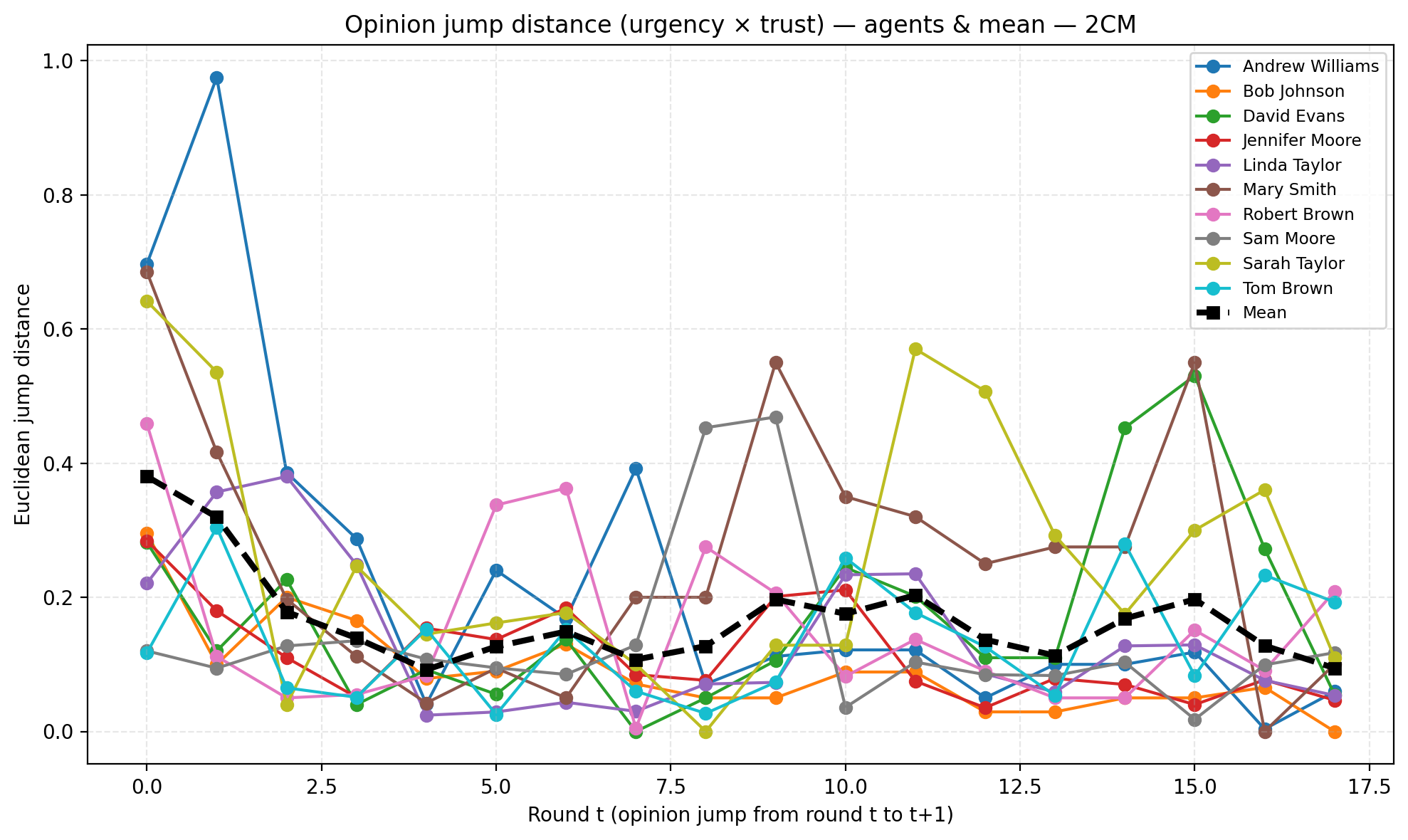}
    \caption{Per‑agent stance distance jumps in the conviction–trust space between rounds $t$ and $t+1$ (colored lines) and the mean jump across agents at each $t$ (black dashed line). For the jumps in conviction–trust space, pairwise $t$‑tests indicate that significant differences obtained only in one comparison, namely, $11\rightarrow 12$ was significantly larger than $12\rightarrow 13$.}
    \label{jump-2CM}
\end{figure}

In this section, we present the results from the experimental condition with two committed minority agents. As in the first run, the second run also displays an increase in the climate urgency conviction during the early rounds, as shown in Figure~\ref{urgency_2CM}. However, the institutional trust in this run converged to around $0$ at earlier stages of the simulation (see Figure~\ref{trust_2CM}). With respect to the stance distance jumps shown in Figure~\ref{jump-2CM}, pairwise $t$-tests indicate that the transition from $11\rightarrow 12$ was significantly larger than $12\rightarrow 13$.

\clearpage
\section{The results for pairwise $t$-tests comparing agent transitions in both experimental conditions}
\label{AppendixE}

In this section, we present the results of the pairwise $t$-tests comparing agent transitions under the experimental conditions with one committed minority and two committed minority members. Under the one committed minority condition, five significant transitions are observed, primarily concentrated in the early interaction rounds. In contrast, in the two committed minority condition, only a single significant comparison emerged near the end of the simulation, specifically, between $11\rightarrow 12$ and $12\rightarrow 13$.

\begin{figure}[htbp]
    \centering   
  \begin{subfigure} [b]{0.49\textwidth}
        \centering
        \includegraphics[width=1.0\linewidth]{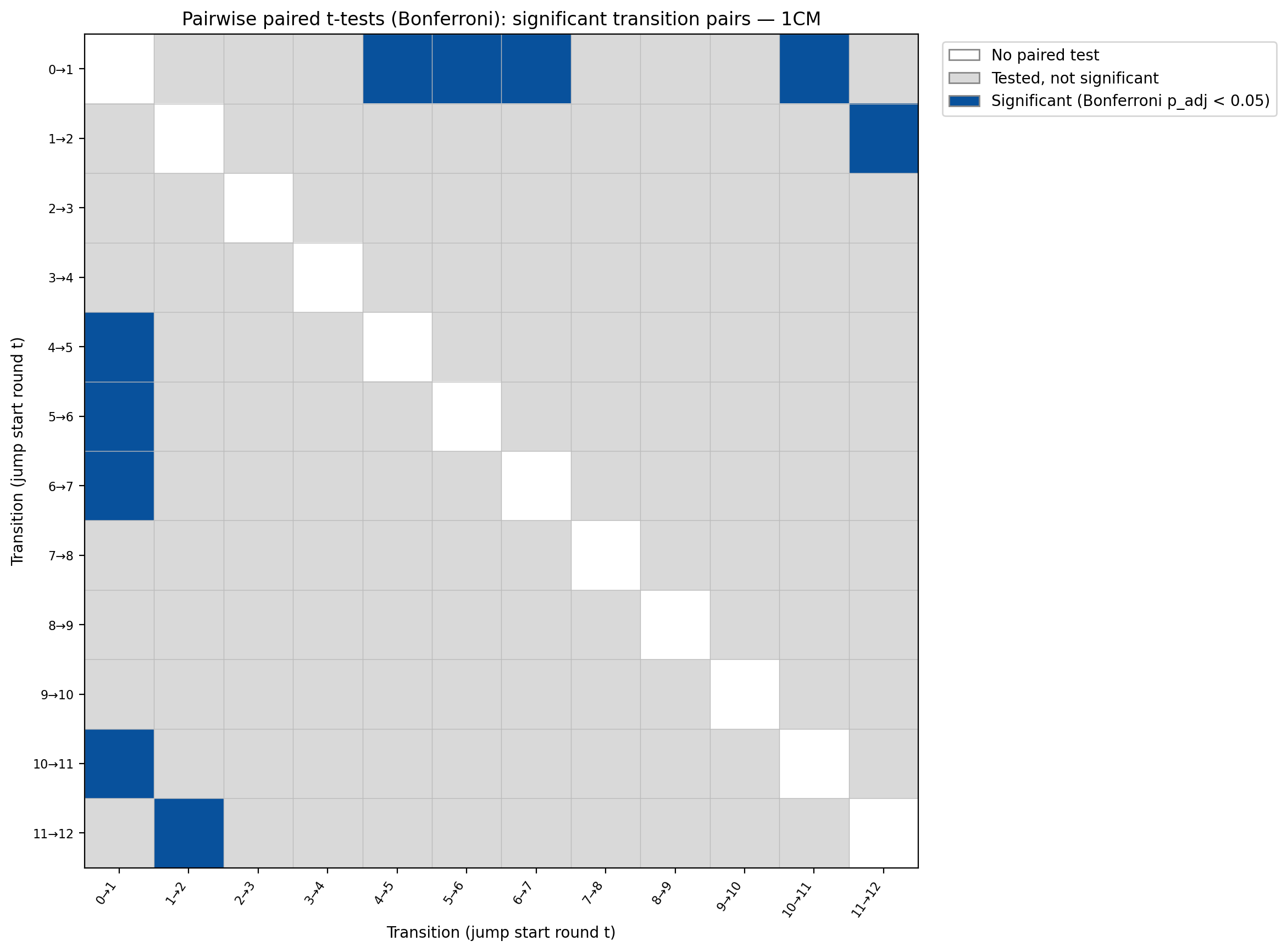}
        \caption{Matrix of pairwise $t$-tests comparing stance-space jump magnitudes between transitions $t\rightarrow t+1$ (rows and columns labeled by jump start round $t$). Experimental condition: one committed minority member.}
        \label{t-test-1CM}
    \end{subfigure}  
    \hfill
    \begin{subfigure} [b]{0.49\textwidth}
        \centering
        \includegraphics[width=1.0\linewidth]{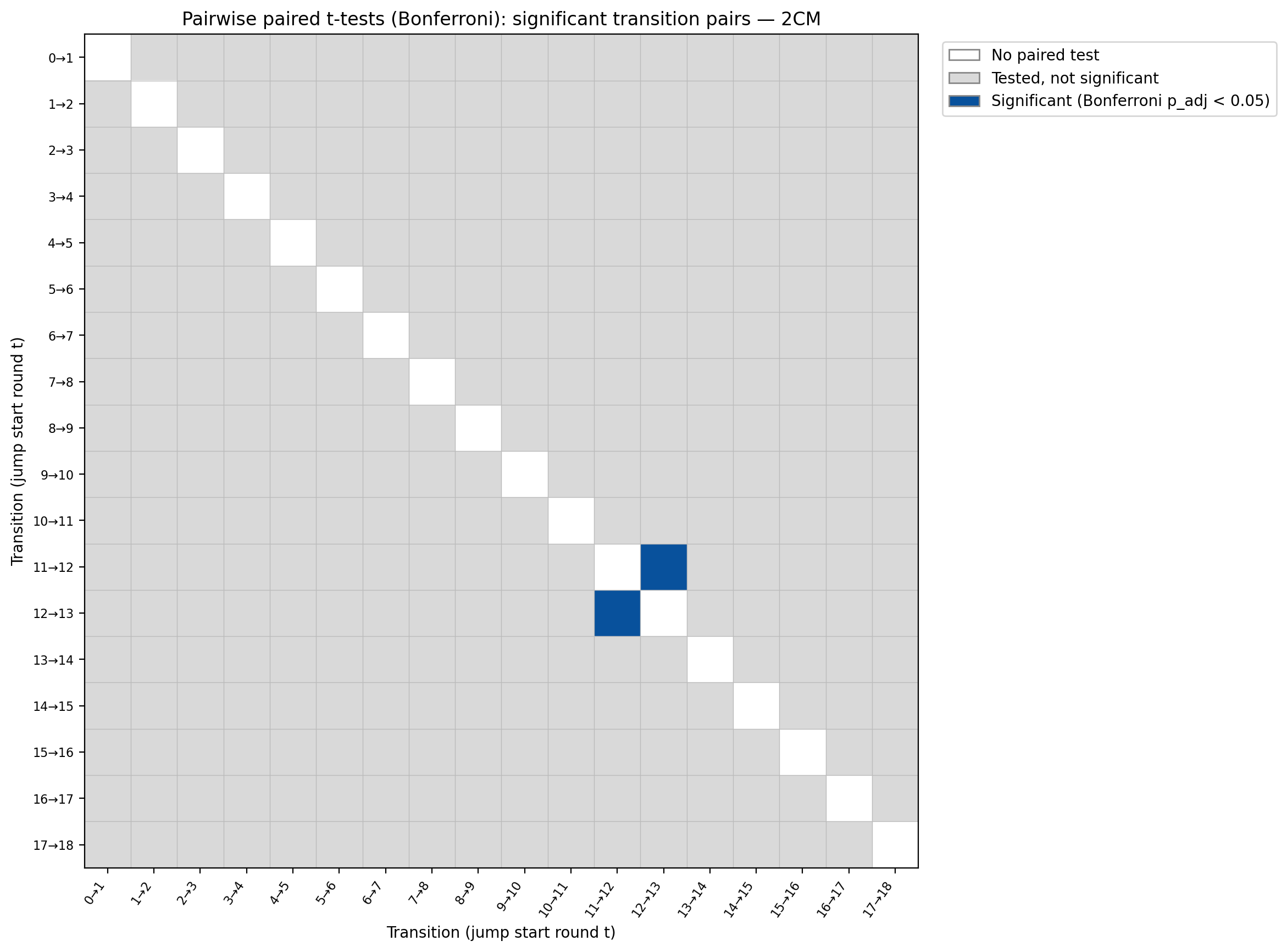}
        \caption{Matrix of pairwise $t$-tests comparing stance-space jump magnitudes between transitions $t\rightarrow t+1$ (rows and columns labeled by jump start round $t$). Experimental condition: two committed minority members.}
        \label{t-test-2CM}
    \end{subfigure}
    \caption{Statistical significance of the jumps in conviction-trust space between rounds $t$ and $t+1$. Dark blue cells indicate pairs for which the difference in mean paired jumps (same agents) is statistically significant after Bonferroni correction for all pairwise tests ($p_{\text{adj}} < \alpha$); light gray indicates a test was run but was not significant; white indicates no paired comparison was available for that cell. The diagonal is empty because a transition is not compared with itself. For the experimental condition of one committed minority (left), significant contrasts are concentrated in comparisons involving early high-movement transitions: $0\rightarrow 1$ is significantly larger than $4\rightarrow 5$, $5\rightarrow 6$, $6\rightarrow 7$, and $10\rightarrow 11$, and $1\rightarrow 2$ is significantly larger than $11\rightarrow 12$. This pattern suggests that agents change stance more strongly in earlier rounds and then stabilize, with only a few later transitions remaining statistically distinguishable from the initial changes. For experimental condition of two committed minority members, the only significant score indicates that transition from $11\rightarrow 12$ was larger than $12\rightarrow 13$.}
    \label{jump_overall-2CM}
\end{figure}

\clearpage
\section{The topic modeling results for the second condition: Two committed minority members}
\label{AppendixF}

In this section, we present the results of topic modeling for the experimental condition with two committed minority agents. The results exhibit dynamics similar to those observed under the one committed minority condition. When the number of topics is constrained to $k=2$, a change of prevalence is observed: topic 0 increases over interaction rounds, while topic 1 declines. However, unlike in the one committed minority condition, the keywords \textit{govern,institut} do not appear among the top 20 words for any tested value of $K \in \{2,3,5\}$.

\begin{figure}[htbp]
    \centering
    \begin{subfigure}[b]{0.32\textwidth}
        \centering
        \includegraphics[width=1.0\linewidth]{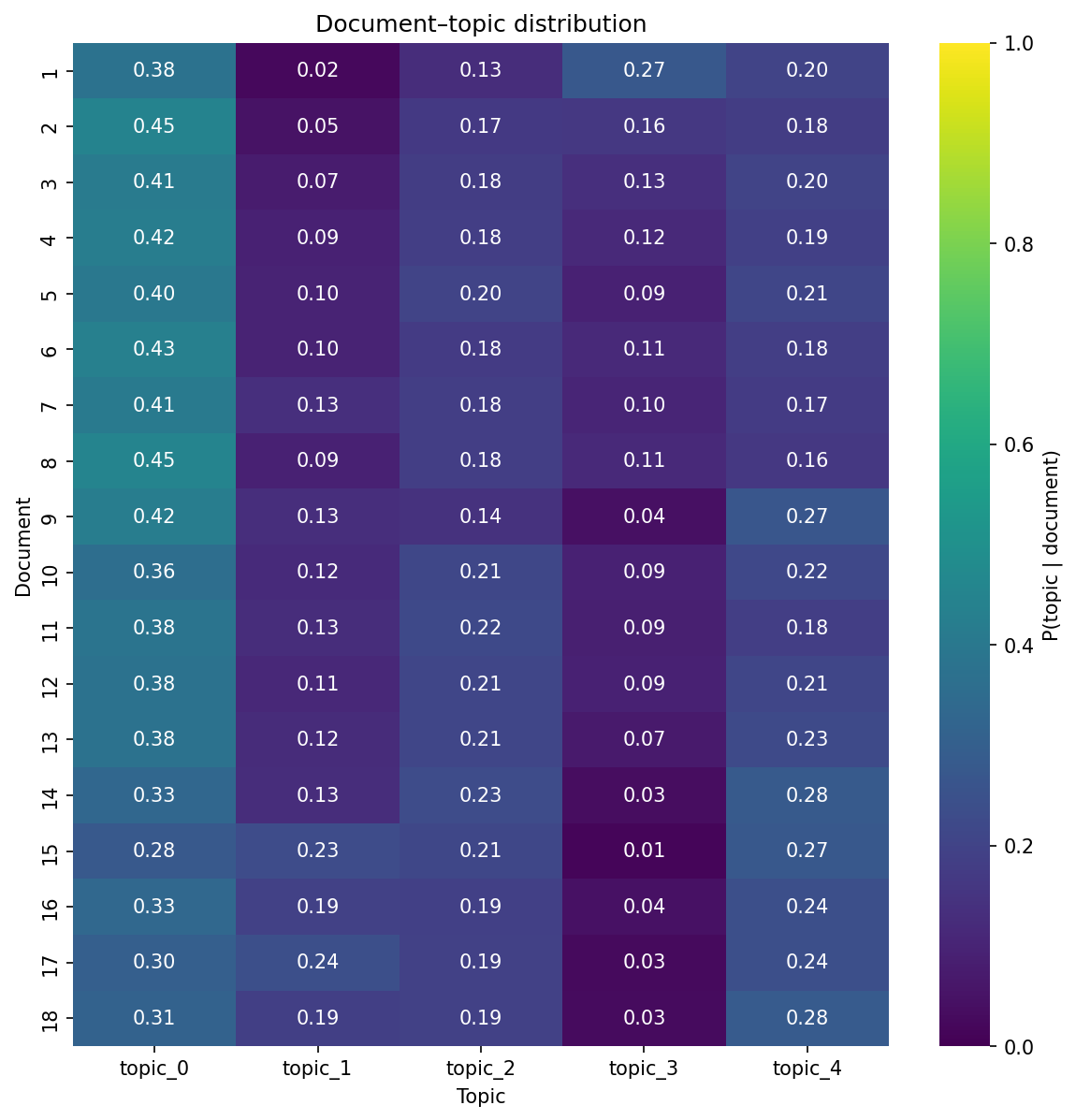}
        \caption{$k$=5}
        \label{fig:placeholder}
    \end{subfigure}
    \hfill
    \begin{subfigure}[b]{0.32\textwidth}
        \centering
        \includegraphics[width=1.0\linewidth]{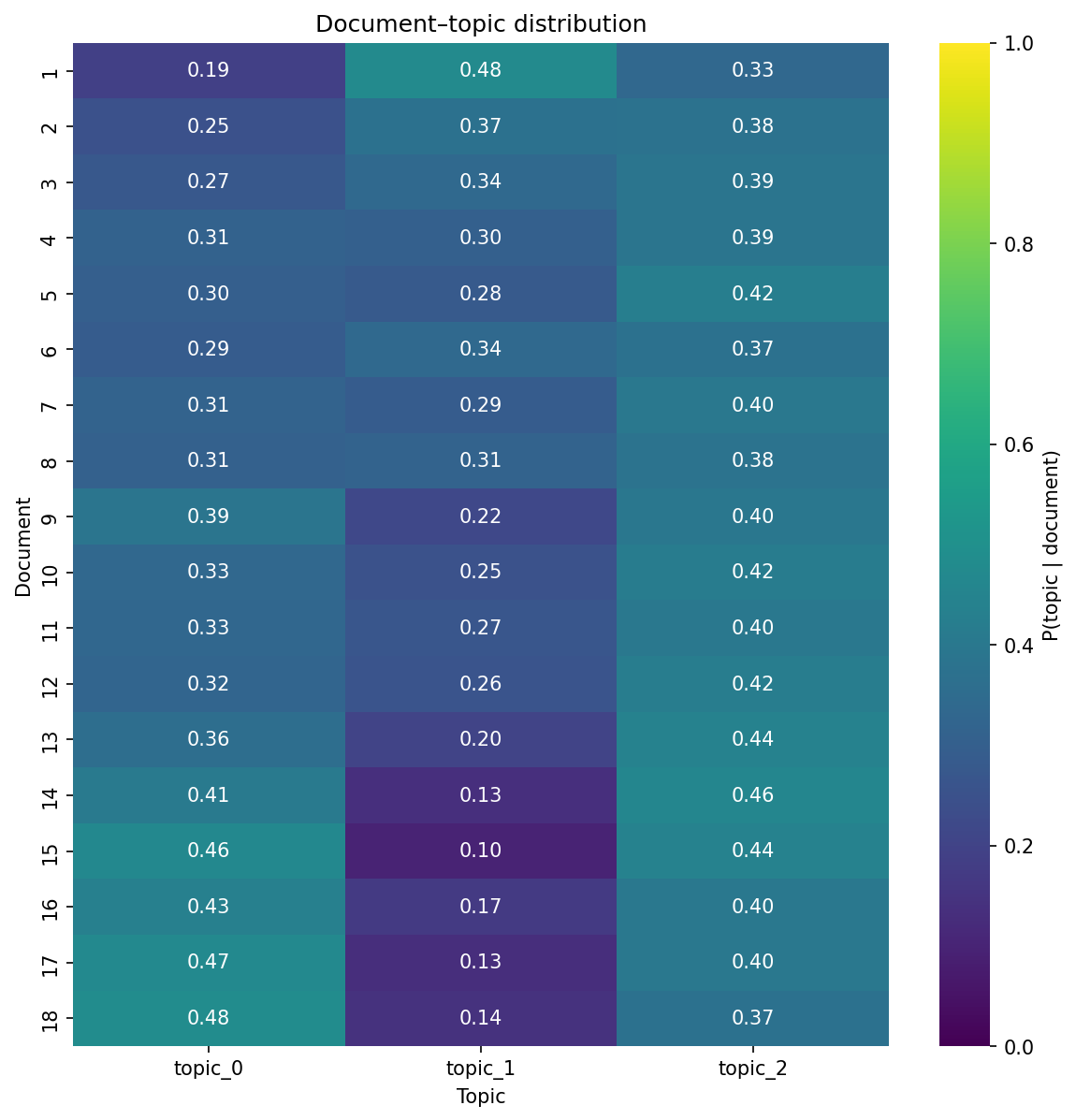}
        \caption{$k$=3}
        \label{fig:placeholder}
    \end{subfigure}
    \hfill
    \begin{subfigure}[b]{0.32\textwidth}
        \centering
        \includegraphics[width=1.0\linewidth]{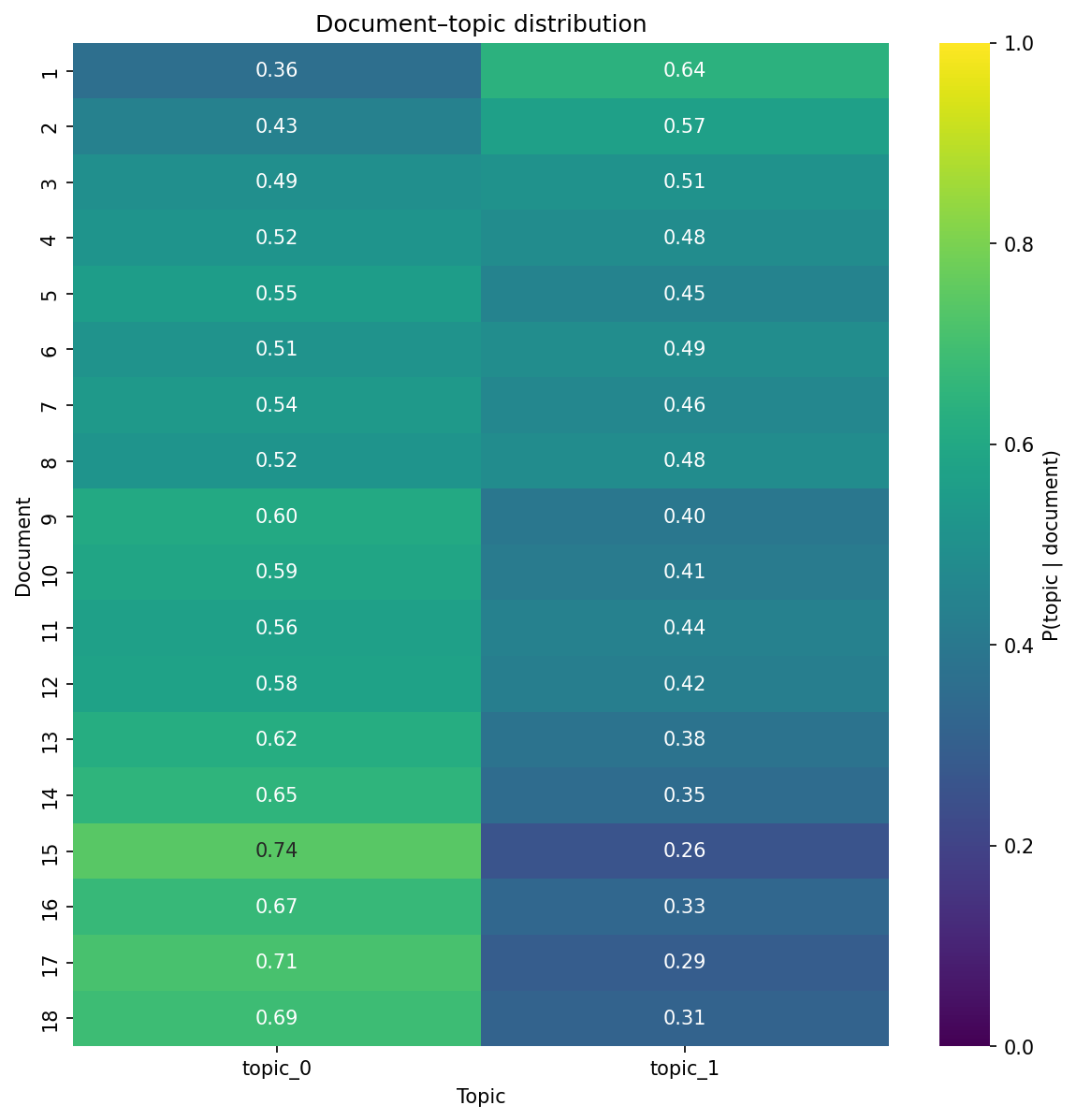}
        \caption{$k$=2}
        \label{fig:placeholder}
    \end{subfigure}
    \caption{Heatmap for the latent topics in the conversation corpora for $k$=2, $k$=3 and $k$=5. Rows correspond to documents defined by interaction rounds, while columns represent topic probabilities. Cell values indicate the probability of a given topic conditioned on a document. The results show that when the model is constrained to $k = 2$, topic 0 exhibits a clear developmental trajectory across rounds, whereas topic 1 shows a corresponding decline in prevalence.}
    \label{LDA_topics_2CM}
\end{figure}

\begin{longtable}{@{}r r >{\raggedright\arraybackslash}p{0.8\linewidth}@{}}
\caption{Top 20 words per topic for $K \in \{2,3,5\}$.} \label{tab:lda-topwords_2CM} \\
\toprule
$K$ & Topic & Top 20 words \\
\midrule
\endfirsthead
\multicolumn{3}{c}{\tablename\ \thetable{} --- \textit{continued}} \\
\toprule
$K$ & Topic & Top 20 words \\
\midrule
\endhead
\midrule
\multicolumn{3}{r}{\textit{Continued on next page}} \\
\endfoot
\bottomrule
\endlastfoot
5 & 0 & communiti stori think climat peopl person help share way like engag workshop brainstorm action let involv inspir chang highlight initi \\
5 & 1 & storytel let artist stori experi definit audienc discuss engag work promot attende theme element narrat feel segment thought list final \\
5 & 2 & idea mayb great make encourag realli particip set session fantast absolut invit interact showcas relat week hey use come specif \\
5 & 3 & action feel discuss organ just individu peopl urgenc definit effect right impact believ talk event exampl garden crucial practic thought \\
5 & 4 & think local realli event creat share reach excit incorpor impact social plan media consid time visual sure school potenti look \\
\midrule
3 & 0 & stori let great storytel experi reach think artist love incorpor connect workshop social sound event way interact collabor set promot \\
3 & 1 & action stori peopl communiti climat help workshop like chang make way feel discuss just success organ initi involv believ set \\
3 & 2 & think realli idea share local communiti engag mayb creat person event encourag brainstorm make particip climat impact excit inspir highlight \\
\midrule
2 & 0 & think realli local event let stori idea engag creat person share great encourag storytel experi particip reach connect set excit \\
2 & 1 & communiti think action stori climat peopl make help share mayb impact realli involv feel inspir workshop chang like idea initi \\
\end{longtable}

\end{appendices}

\end{document}